\PassOptionsToPackage{unicode}{hyperref}
\PassOptionsToPackage{hyphens}{url}
\PassOptionsToPackage{dvipsnames,svgnames,x11names,table}{xcolor}
\documentclass[12pt]{article}
\usepackage{pdfpages}
\usepackage{makecell}
\usepackage{multirow}
\usepackage{longtable}
\usepackage{ragged2e}
\usepackage{booktabs}
\usepackage{amsmath,amssymb}
\usepackage{setspace}
\usepackage{xstring}
\usepackage{iftex}
\usepackage{verbatim}
\ifPDFTeX
  \usepackage[T1]{fontenc}
  \usepackage[utf8]{inputenc}
  \usepackage{textcomp} 
\else 
  \usepackage{unicode-math}
  \defaultfontfeatures{Scale=MatchLowercase}
  \defaultfontfeatures[\rmfamily]{Ligatures=TeX,Scale=1}
\fi
\usepackage{lmodern}
\ifPDFTeX\else  
\fi
\IfFileExists{upquote.sty}{\usepackage{upquote}}{}
\IfFileExists{microtype.sty}{
  \usepackage[]{microtype}
  \UseMicrotypeSet[protrusion]{basicmath} 
}{}
\makeatletter
\@ifundefined{KOMAClassName}{
  \IfFileExists{parskip.sty}{%
    \usepackage{parskip}
  }{
    \setlength{\parindent}{0pt}
    \setlength{\parskip}{6pt plus 2pt minus 1pt}}
}{
  \KOMAoptions{parskip=half}}
\makeatother
\usepackage{xcolor}
\makeatletter
\ifx\paragraph\undefined\else
  \let\oldparagraph\paragraph
  \renewcommand{\paragraph}{
    \@ifstar
      \xxxParagraphStar
      \xxxParagraphNoStar
  }
  \newcommand{\xxxParagraphStar}[1]{\oldparagraph*{#1}\mbox{}}
  \newcommand{\xxxParagraphNoStar}[1]{\oldparagraph{#1}\mbox{}}
\fi
\ifx\subparagraph\undefined\else
  \let\oldsubparagraph\subparagraph
  \renewcommand{\subparagraph}{
    \@ifstar
      \xxxSubParagraphStar
      \xxxSubParagraphNoStar
  }
  \newcommand{\xxxSubParagraphStar}[1]{\oldsubparagraph*{#1}\mbox{}}
  \newcommand{\xxxSubParagraphNoStar}[1]{\oldsubparagraph{#1}\mbox{}}
\fi
\makeatother

\usepackage{longtable,booktabs,array}
\usepackage{calc} 
\usepackage{etoolbox}
\makeatletter
\patchcmd\longtable{\par}{\if@noskipsec\mbox{}\fi\par}{}{}
\makeatother
\IfFileExists{footnotehyper.sty}{\usepackage{footnotehyper}}{\usepackage{footnote}}
\makesavenoteenv{longtable}
\usepackage{graphicx}
\makeatletter
\def\maxwidth{\ifdim\Gin@nat@width>\linewidth\linewidth\else\Gin@nat@width\fi}
\def\maxheight{\ifdim\Gin@nat@height>\textheight\textheight\else\Gin@nat@height\fi}
\makeatother
\setkeys{Gin}{width=\maxwidth,height=\maxheight,keepaspectratio}
\makeatletter
\def\fps@figure{htbp}
\makeatother

\makeatletter
\@ifpackageloaded{caption}{}{\usepackage{caption}}
\AtBeginDocument{%
\ifdefined\contentsname
  \renewcommand*\contentsname{Table of contents}
\else
  \newcommand\contentsname{Table of contents}
\fi
\ifdefined\listfigurename
  \renewcommand*\listfigurename{List of Figures}
\else
  \newcommand\listfigurename{List of Figures}
\fi
\ifdefined\listtablename
  \renewcommand*\listtablename{List of Tables}
\else
  \newcommand\listtablename{List of Tables}
\fi
\ifdefined\figurename
  \renewcommand*\figurename{Figure}
\else
  \newcommand\figurename{Figure}
\fi
\ifdefined\tablename
  \renewcommand*\tablename{Table}
\else
  \newcommand\tablename{Table}
\fi
}
\@ifpackageloaded{float}{}{\usepackage{float}}
\floatstyle{ruled}
\@ifundefined{c@chapter}{\newfloat{codelisting}{h}{lop}}{\newfloat{codelisting}{h}{lop}[chapter]}
\floatname{codelisting}{Listing}

\makeatother
\makeatletter
\@ifpackageloaded{caption}{}{\usepackage{caption}}
\@ifpackageloaded{subcaption}{}{\usepackage{subcaption}}
\makeatother

\ifLuaTeX
  \usepackage{selnolig}  
\fi
\usepackage[]{natbib}
\usepackage{bookmark}

\IfFileExists{xurl.sty}{\usepackage{xurl}}{} 
\hypersetup{
  pdftitle={Title},
  pdfauthor={Author 1; Author 2},
  pdfkeywords={3 to 6 keywords, that do not appear in the title},
  colorlinks=true,
  linkcolor={blue},
  filecolor={Maroon},
  citecolor={Blue},
  urlcolor={Blue},
  pdfcreator={LaTeX via pandoc}}

\newcommand{\anon}{1}

\newcommand{\versionphrase}{%
  \ifnum\anon=0
    a private research university%
  \else
    Carnegie Mellon University%
  \fi
}

\newcommand{\coursetoggle}{%
  \ifnum\anon=0
    \textit{Statistics 102}%
  \else
    \textit{Methods for Statistics \& Data Science}%
  \fi
}

\begin{document}

\definecolor{lightgray}{gray}{0.9}
\def\spacingset#1{\renewcommand{\baselinestretch}%
{#1}\small\normalsize} \spacingset{1}

\definecolor{llmcolor}{HTML}{56B4E9}
\colorlet{llmcolor}{llmcolor!20}

\definecolor{studentcolor}{HTML}{000000}
\colorlet{studentcolor}{studentcolor!20}

\definecolor{expertcolor}{HTML}{E69F00}
\colorlet{expertcolor}{expertcolor!30}

\newcommand{\llmrow}{\rowcolor{llmcolor}}
\newcommand{\expertrow}{\rowcolor{expertcolor}}
\newcommand{\studentrow}{\rowcolor{studentcolor}}


\if1\anon
{
  \title{\bf Analyzing Students’ Statistics Writing Before and After the Emergence of Large Language Models}
  \author{Sara Colando\thanks{These authors contributed equally to this work.}\hspace{.2cm}\\
    Department of Statistics \& Data Science, Carnegie Mellon University\\
    Erin Franke\footnotemark[1]\hspace{.2cm}\\
     Department of Statistics \& Data Science, Carnegie Mellon University\\
    Gordon Weinberg\\
     Department of Statistics \& Data Science, Carnegie Mellon University\\
    and \\
    Alex Reinhart\\
     Department of Statistics \& Data Science, Carnegie Mellon University}
  \maketitle
} \fi

\if0\anon
{
  \bigskip
  \bigskip
  \bigskip
  \begin{center}
    {\LARGE\bf \bf Analyzing Students’ Statistics Writing Before and After the Emergence of Large Language Models}
\end{center}
  \medskip
} \fi

\bigskip
\begin{abstract}
The ability to communicate statistical results to domain experts and stakeholders is an important goal
of the undergraduate statistics and data science curriculum. However, as large language models (LLMs) have become more accessible, a major concern is that students are offloading important cognitive tasks to generative AI. Using a corpus of over 1,600 undergraduate students' data analysis reports from 2021 to 2025, we show how students’ writing style and verb usage have become more similar to that of LLMs. This shift is most pronounced in the first and
fifth quintiles of students’ reports, which roughly map onto the introduction and conclusion sections, respectively. At the same time, we demonstrate that students’ writing style has become more similar to that of statistics experts with the addition of LLMs. We end by discussing the implications of our findings for statistics and data science educators. In particular, we propose alternative modes of assessment that still emphasize statistical thinking, such as targeted writing assignments for structuring a report introduction.

\end{abstract}

\noindent%
\vfill

\newpage
\spacingset{1.8} 

\section{Introduction}\label{sec-intro}

Project-based learning, a pedagogical approach in which students complete course projects that address real-world problems, has been widely adopted in statistics and data science education since the 1990s \citep{AlLabadi2025}. In statistics and data science courses, project-based assessments can help students to apply course content to meaningful contexts while simultaneously developing end-to-end data analysis, communication, and collaboration skills (e.g., see \citealp{Ritter2024}).

Writing assignments are a common form of project-based assessment in statistics and data science curricula and serve multiple pedagogical purposes. For one, the writing process encourages students to organize and articulate their thoughts clearly, which may help them develop a deeper conceptual understanding of course material \citep{Woodard2020}. These \textit{writing-to-learn} assignments are particularly useful in introductory statistics and data science courses, where students are exposed to several fundamental concepts, such as hypothesis testing, data visualization, statistical modeling, and interpreting statistical results in context. On the other hand, writing assignments can also help students improve their ability to communicate complex statistical results to non-expert audiences, such as stakeholders or domain experts  \citep{Woodard2020}. In these \textit{writing-in-the-disciplines} assignments, students tailor their communication of statistical results to their intended audience's background, which is an important skill for modern statistics and data science practitioners. Given their purpose, writing-in-the-disciplines assignments are often used in advanced undergraduate courses, like capstone courses \citep{Spurrier2001, Martonosi2016}.

Through writing-to-learn and writing-in-the-disciplines assignments, students can evolve from novice to expert writers in statistics and data science. According to \cite{Kellogg2008}, writing skill development can be divided into three stages: knowledge-telling, knowledge-transforming, and knowledge-crafting. In the knowledge-telling stage, students primarily use writing to state what they already know \citep{Kellogg2008}. Students build on the knowledge-telling stage with the knowledge-transforming stage, in which they use writing to think through a topic, refine their ideas, and construct new understanding \citep{Kellogg2008}. This stage of the writing process closely aligns with the primary goals of writing-to-learn assignments, in which students use writing to deepen their understanding of statistical concepts. Finally, in the knowledge-crafting stage, students actively consider their audience during the writing process by imagining the reader's interpretation of their text and adapting it accordingly \citep{Kellogg2008}. In the context of statistics and data science, reaching the knowledge-crafting stage requires that students tailor their communication of statistical methods and results to their intended audience's background. From this perspective, a central goal of writing-in-the-disciplines assignments is to help students reach the knowledge-crafting stage, which signals expertise in statistics and data science writing. 

With the rise of large language models (LLMs), such as ChatGPT, Gemini, and Claude, a major concern is that students are offloading important cognitive tasks that writing assignments and other traditional project-based learning assessments were originally intended to invoke. 
For instance, if a student feeds their linear regression results to an LLM and asks it to interpret the coefficients, they overlook the important skill of interpreting statistical results in context. In turn, they may fail to achieve the first goal of the GAISE 2016 College Guidelines that ``students should become critical consumers of statistically-based results" \citep{gaise}. Similarly, asking an LLM to write the limitations of a data analysis may impede students from carefully thinking through relevant ethical considerations or how additional data sources, covariates, or context could improve their analysis. This neglects the second goal in the GAISE 2016 College Guidelines, which requires students to clearly communicate the results and limitations of their data analysis \citep{gaise}. In terms of writing development, heavy reliance on generative AI may compromise students' writing-skill trajectories \citep{DeLuca2025}. Indeed, \cite{Georgiou2025} finds that participants who did not use ChatGPT in a writing assignment were more likely to report engaging in deep processing, exerting more mental effort, and having stronger sustained attention during the writing task. Thus, students who repeatedly offload writing tasks to LLMs may remain in the knowledge-telling phase, where they simply reiterate their (or the LLM's) thoughts in written form. These students might struggle to communicate their findings on the spot in their future careers, produce statistical results in novel, non-classroom settings, and lack the skills necessary to be critical consumers of the data and statistical claims presented to them.

In this paper, we evaluate how students' statistics writing has changed, in terms of both writing style and verb usage, since the introduction of LLMs. To do so, we compare a corpus of data analysis reports written by introductory statistics students from 2021 to 2025 to a corpus of academic text generated by four LLMs: GPT-4o, GPT-5 Mini, Gemini Flash, and Claude Haiku.  
We also compare the writing style in students' reports to that of expert statistics writing using a corpus of peer-reviewed applied statistics publications from \textit{CHANCE} magazine, \textit{The American Statistician}, and the \textit{Journal of the American Statistical Association} in order to assess whether students' writing has become more expert-like in recent years. We discuss the implications of our findings for statistics and data science educators and propose alternative project-based assessment methods, like targeted writing assessments, based on our results.

\section{Background}\label{sec-back}

\subsection{Human versus LLM Writing Style}

Previous research has identified several differences between human and LLM writing styles. A consistent finding is that LLM-generated text exhibits greater structural complexity than human-authored text. In particular, LLM-generated text often includes more nominalizations --- nouns formed from verbs or adjectives (e.g., \textit{operation} from \textit{operate}, \textit{commitment} from \textit{commit}, \textit{flexibility} from \textit{flexible}) --- than human writing \citep{Herbold2023, Mizumoto2024, Jiang2025, Reinhart2025, DeLuca2025}. LLM-generated text also tends to use more present participial clauses (e.g., \textit{having a dog}, \textit{writing a letter}) \citep{Reinhart2025}, longer words \citep{DeLuca2025}, more passive voice \citep{Reinhart2025}, greater lexical diversity \citep{Mizumoto2024}, and have a more positive sentiment \citep{Mak2025} than human-authored text. In contrast, human-authored writing tends to contain more epistemic constructions than LLM-generated text \citep{Herbold2023, Mizumoto2024, Jiang2025}. Epistemic constructions, like model verbs (e.g., \textit{might}, \textit{can}, \textit{should}, \textit{must}), allow the writer to express their uncertainty or commitment to the content of their writing. 

There is also evidence that LLMs tend to favor specific words more than human writers. \cite{Reinhart2025} found that OpenAI's GPT-4 and 4o Mini use words like \textit{camaraderie, palpable, tapestry}, and \textit{intricate} at more than 100 times the rate of human writers. Moreover, \cite{Kobak2025} examined year-to-year changes in word frequencies across more than fifteen million English-language biomedical abstracts indexed by PubMed from 2010--2024. Several words were used substantially more often in abstracts from 2024, including common words such as \textit{potential}, \textit{findings}, and \textit{crucial}, as well as less common words such as \textit{delves}, \textit{underscores}, and \textit{showcasing} \citep{Kobak2025}.

\subsection{Student Writing in the Age of Generative AI} 

At the same time, several recent studies have examined how student writing has changed with the introduction of generative AI. \cite{Mak2025} examined the changes in lexical features, nominalizations, syntactic complexity, and readability of 4,820 reports written by undergraduate psychology majors in the United Kingdom from 2016--2025. They found that the use of eight words which had been previously linked to ChatGPT-generated writing (\textit{crucial, comprehensive, intricate, pivotal, delve, underscore, utilise/ze, align}, and their inflection forms) increased sharply in student reports between 2023 and 2024 \citep{Mak2025}. The use of these eight words in student reports then declined in 2025, possibly because students were trying to mask AI involvement \citep{Mak2025}.

Similar patterns have been observed in student writing from online learning platforms. \cite{mooc} analyzed essay submissions from 3,582 participants enrolled in a free Massive Open Online Course on AI ethics. They compared writing submissions from before ChatGPT's release (late 2020 to November 2022) to those written one year following its release (December 2023 or later). The average response length of students' submissions significantly increased between the two time periods, with a sharp jump in March 2023 \citep{mooc}. There was also a shift in students' response topics, despite the writing prompt being identical across the two periods \citep{mooc}. Similar to \cite{Mak2025}, they also found that ChatGPT-associated words such as \textit{delve, foster,} and \textit{crucial} were used more frequently in students' submissions after ChatGPT was released \citep{mooc}. 

This article extends previous work on how student writing has changed in the age of generative AI in three ways. First, we focus on writing in statistics and data science courses, where writing is used throughout the curriculum via writing-to-learn and writing-in-the-disciplines assignments. Second, we analyze a more comprehensive set of linguistic features to characterize writing style in student reports and LLM-generated text. Finally, we compare students' writing styles with those of domain experts to assess whether student writing has become more expert-like since the introduction of LLMs.

\section{Corpus Data}
\label{sec-corpora}

We use three corpora to assess the trajectory of students' statistics writing since the emergence of LLMs. Our ``student corpus" consists of introductory statistics students' data analysis reports from 2021--2025 (Section \ref{subsec-student-corpus}). Our ``LLM corpus" contains academic writing generated by GPT, Gemini, and Claude LLMs (Section \ref{subsec-LLM-corpus}). Finally, our ``expert corpus" contains peer-reviewed applied statistics articles that were published in \textit{Chance} magazine, \textit{The American Statistician}, and \textit{Journal of the American Statistical Association} before the release of LLMs (Section \ref{subsec-expert-corpus}).

\subsection{Student Corpus}
\label{subsec-student-corpus}

Our student corpus consists of undergraduate student data analysis reports from \textit{36-202:} \coursetoggle{}{} at \versionphrase. \coursetoggle{}{} is an introductory undergraduate statistics course, typically taken during students' first or second year. It is a required course for Statistics \& Data Science majors, and a recommended course for several other majors, including Business Administration, Information Systems, and Social and Decision Sciences. Students must either take an introductory statistics course at the university or score a 5 on the AP Statistics exam to enroll in the course. In the course, students are introduced to several fundamental statistics topics, such as simple and multiple regression, basic analysis of variance (ANOVA), logistic regression, and data mining methods, such as classification and clustering. A central objective of the course is that students gain experience applying these statistical methods to real-world data through coding assignments and data analysis reports written in R Markdown. As the university requires students to fulfill a writing requirement during their first semester, all students enrolled in \coursetoggle{}{} have either already completed a college-level writing course or are concurrently enrolled in one. 

Reports from the first data analysis project in the course, which is due halfway through the semester, are included in our student corpus. In this data analysis project, students build and justify a multiple linear regression model using the regression topics covered in lecture. They are also asked to predict the response variable for provided values of potential explanatory variables. Students are given a choice among four datasets and corresponding prompts to work from, with topics including: bike-sharing systems, New York City housing, court cases, and sharing articles on social media. These options are consistent across all semesters included in the student corpus. A sample prompt and project instructions are available in the supplementary materials. 

The students' reports are expected to follow a modified IDMRaD structure, which includes introduction, exploratory data analysis, modeling, prediction, and discussion sections. The instructor provides students with an example report prior to the assignment in order to show how the report should be formatted. The report is graded out of 100 points, with a three-point penalty applied for every sentence or phrase copied verbatim or nearly verbatim from the exemplar. While no explicit policy on generative AI usage is included in the project instructions, the instructions do state that ``the work and words you submit must be your own". Students submit their final data analysis report in both R Markdown and PDF formats, which are then graded by undergraduate- and master's-level teaching assistants. 

Table \ref{tab:student-corpus} displays the breakdown of the student corpus by semester. In total, the corpus contains 1,619 data analysis reports written between the Spring 2021 and Fall 2025 semesters. The only exception is the Fall 2023 semester, when the instructor neglected to collect the R Markdown documents. The collection protocol was approved by the university's Institutional Review Board (STUDY2020\_00000322). Students were given the option to opt out of having their report included in research at the time of their submission, with no penalty on their grade. Before analysis, all reports were anonymized by removing the headers from the R Markdown documents. 

\begin{table}

\centering

\rowcolors{3}{}{lightgray}

\begin{tabular}{p{3.5cm}rrr}

\toprule

Semester & Submitted reports & Collected reports & Collected report tokens \\

\midrule

\multicolumn{4}{l}{\underline{\textit{Pre-LLM}}} \\

\quad Spring 2021 &  222 &  206 & 350,999 \\

\quad Fall 2021  & 179 & 145 & 261,004 \\

\quad Spring 2022  & 194 & 170 & 308,466 \\

\quad Fall 2022 & 194 & 170 & 313,273 \\

\midrule

\multicolumn{4}{l}{\underline{\textit{Transition Year}}} \\

\quad Spring 2023 & 192 & 170 & 301,208 \\

\quad Fall 2023  & 192 & -- & -- \\

\midrule

\multicolumn{4}{l}{\underline{\textit{Since-LLM}}} \\

\quad Spring 2024 & 202 & 178 & 343,533 \\

\quad Fall 2024 & 205 & 187 & 357,165 \\

\quad Spring 2025 & 218 & 187 & 326,012 \\

\quad Fall 2025 & 230 & 206 & 387,493 \\

\bottomrule

\end{tabular}
\caption{Composition of the student corpus. All texts are written by undergraduate students in an introductory statistics and data science course at Carnegie Mellon University. The difference between submitted reports and collected reports constitutes students who opted out of having their reports used for research. We consider reports from Spring 2021 to Fall 2022 to be \textit{pre-LLM}, and reports from Spring 2024 to Fall 2025 to be \textit{since-LLM}. Note that we are missing reports from Fall 2023 because the instructor neglected to collect the R Markdown documents that semester.}
\label{tab:student-corpus}
\end{table}

As our goal is to examine how student writing has changed with the widespread adoption of LLMs, we categorized semesters into \textit{pre-LLM}, \textit{transition year}, and \textit{since-LLM} periods. We designated the \textit{pre-LLM} period to include student reports from 2021--2022, since OpenAI first launched ChatGPT on November 30, 2022 \citep{ChatGPT-first}. Since generative AI tools gained popularity and widespread adoption by the public during 2023, we denote 2023 as a \textit{transition year}. By Spring 2024, several additional LLMs had become widely accessible, including Anthropic's Claude, released March 2023 \citep{Claude2023}, and Google's Gemini, released December 2023 \citep{Gemini2023}. As a result, we regard student reports coming from 2024--2025 as \textit{since-LLM} reports. 

\subsection{LLM Corpus}
\label{subsec-LLM-corpus}

Our LLM corpus is a subset of the Human-AI Parallel English (HAP-E) academic corpus created by \cite{Reinhart2025}. The HAP-E academic corpus was created by sampling 2,000 articles from a corpus of over 40,000 open-access papers published by Elsevier. The first 1,000 words of each sampled article were selected and split into two (roughly) 500-word chunks, with the breaking point at a sentence boundary \citep{Reinhart2025}. The LLMs were given the first 500-word chunk of each sampled article and prompted to generate the next 500 words in the same style, tone, and diction. All LLM nonresponses and generated texts shorter than 100 words were excluded from the corpus. If an article corresponded to an LLM nonresponse, \cite{Reinhart2025} also removed the other LLM-generated texts associated with this article in order to facilitate comparison between different LLMs. After filtering, the HAP-E academic corpus contained LLM-generated text associated with 1,227 open-access articles. 

For our analysis, we used academic texts generated by four LLMs: GPT-4o \citep{ChatGPT4o}, GPT-5 Mini \citep{ChatGPT5}, Gemini 2.5 Flash \citep{GeminiFlash}, and Claude Haiku 4.5 \citep{ClaudeHaiku4.5}. We included these LLMs because each was easily accessible to undergraduate students at \versionphrase{}{} at some point between the Spring 2023 and Fall 2025 semesters. Additionally, OpenAI's ChatGPT, Google's Gemini, and Anthropic's Claude models are among the most popular LLMs for coding and writing tasks based on \href{https://arena.ai/leaderboard}{arena.ai} leaderboard rankings from February 11, 2026. Hence, we expected these four LLMs to be representative of those that undergraduate students in \coursetoggle{}{} might consult when writing their reports. We provide the release date, collection date, number of texts, and total number of tokens for each of these four LLMs in Table \ref{tab:hape-corpus}.

\begin{table}
\centering
\rowcolors{1}{}{lightgray}
\begin{tabular}{lllcc}
\toprule
Large Language Model & Release Date & Collection Date & Texts & Tokens \\
\midrule
GPT-4o & August 6, 2024 & September 9, 2024 & 1,227 & 703,991	\\

GPT-5 Mini & August 7, 2025  & August 10, 2025 & 1,227 & 712,061	 \\

Gemini 2.5 Flash & June 17, 2025 & January 23, 2026 & 1,227 & 574,559 \\

Claude Haiku 4.5 & October 1, 2025 & December 23, 2025 & 1,227 & 606,200 \\
\bottomrule
\end{tabular}
\caption{Composition of the LLM corpus, which is a subset of the HAP-E academic corpus created by \cite{Reinhart2025}. GPT and Claude release dates are based on the model version date, not the date that the model was announced.}
\label{tab:hape-corpus}
\end{table}

\subsection{Expert Corpus}
\label{subsec-expert-corpus}

Finally, our expert corpus consists of peer-reviewed applied statistics articles from \textit{CHANCE} magazine, the \textit{Journal of the American Statistical Association} (Applications and Case Studies), and \textit{The American Statistician} (Statistical Practice). We only included articles published between 2018 and 2022 to ensure they did not contain any LLM-generated text. This resulted in a total of 287 articles. We downloaded the 287 articles as electronic publications (EPUBs) and converted them into text files. All headers, figures, and in-line math in the text files were replaced with flags. We defined in-line math as either figures with no caption (as in-line math was represented as GIFs in some articles) or proper math objects recognized by the Pandoc document parser. Next, we computed the math-to-word ratio in each article section by dividing the number of math flags by the number of words. We removed article sections with a math-to-word ratio greater than 0.01 so that documents in the expert corpus more closely resembled exemplar versions of the student reports, which are not expected to contain much math. We excluded articles with fewer than 500 words remaining so that we had sufficient text to gauge each article's writing style. In practice, this meant that statistical theory and probability articles, as well as simulation studies, were usually excluded from the expert corpus. After preprocessing, our expert corpus consists of 278 articles. Table \ref{tab:expert-corpus} shows the publication period, number of articles, and total tokens for each of the three journals.

\begin{table}
\centering
\rowcolors{1}{}{lightgray}
\begin{spacing}{1.25}
\renewcommand{\arraystretch}{1.2}
\begin{tabular}{p{9cm}ccc}
\toprule
Journal (article type, if applicable) & Publication period & Texts & Tokens \\
\toprule
\textit{CHANCE} & 2018--2022 & 98 & 315,458 \\
\textit{The American Statistician} (Statistical Practice) & 2019--2022 & 38 & 131,874 \\
\textit{Journal of the American Statistical Association} (Applications and Case Studies) & 2019--2022 & 142 &  565,279	\\
\bottomrule
\end{tabular}
\end{spacing}
\caption{Composition of the expert corpus. Note that the collection years for the \textit{The American Statistician} and the \textit{Journal of the American Statistical Association} articles started in 2019, as this was when EPUB versions of articles were first available for download.}
\label{tab:expert-corpus}
\end{table}

\section{Methods}

\subsection{Text Tagging}
\label{subsec:text-tagging}

We tokenized and tagged each document in the three corpora using the \texttt{spacyr} R package  \citep{spacy}. This provided us with the lemma and part-of-speech tag (e.g., \textit{noun, verb, auxiliary, punctuation, pronoun}) for each token in a document. The lemma is the root, or dictionary form, of a word \citep{Zgusta2006}; for example, the lemma of \textit{found} is \textit{find}, and the lemma of \textit{are} is \textit{be}.

We then extracted the Biber feature rates (per 1,000 tokens) from our tokenized and tagged corpora with the \texttt{pseudobibeR} R package \citep{pseudobibeR}. Biber features are a set of 67 lexicogrammatical and functional features of writing, such as the mean word length, tense and aspect markers (e.g., past tense usage), pronoun markers (e.g., first person pronoun usage), passives (e.g., passive voice usage), and lexical classes (e.g., downtoner usage like \textit{barely}, \textit{nearly}, and \textit{slightly}) \citep{Biber1985}. They have been effective for distinguishing between expert and novice writing \citep{Biber2002, Hardy2013, Staples2016}, between human-authored and LLM-generated text \citep{Reinhart2025}, and between students' statistics writing, expert writing, and LLM-generated text \citep{DeLuca2025}. We provide a table of all 67 Biber features, along with descriptions and examples from the student corpus, in the Supplementary Materials (Table \ref{tab:biber-features}). We did not include the type-token ratio (i.e., the ratio of unique words to total words) in our analysis, because it is strongly correlated with document length \citep{Koizumi2012}, which varies systematically across our corpora.   

\subsection{Text Analysis}


\subsubsection{Comparing Writing Style}
\label{subsec:writing-style}

To compare writing style across the corpora, we fit two Linear Discriminant Analysis (LDA) models, which we refer to as the ``overall LDA model" and the ``quintile LDA model". LDA is a classification and dimension reduction technique that maximizes the ratio of between-source to within-source variance among the reference sources \citep{Hastie2009}. The LDA dimensions are linear combinations of the original feature space and can be used to estimate the posterior probability that an observation belongs to each reference source. LDA is often effective for distinguishing between text sources. Most relevant to our application, \cite{DeLuca2025} used LDA to separate machine-generated, expert, and undergraduate students' statistics introductions. Yet, LDA has been used for other text classification tasks, such as for distinguishing between online genre categories \citep{Biber2016} and company email types \citep{Brown2021}.

We trained both the overall LDA model and the quintile LDA model using jointly standardized Biber feature rates from (i) \textit{pre-LLM} student reports, (ii) LLM-generated text, and (iii) expert writing. Thus, both LDA models projected the original 66-dimensional space of standardized Biber feature rates onto a two-dimensional subspace, spanned by a linear combination of the original features.

\textbf{Overall LDA Model:} For the overall LDA model, we first jointly standardized the Biber feature rates across four text sources: \textit{pre-LLM} student reports, \textit{since-LLM} student reports, LLM-generated text, and expert writing. We then randomly assigned 75\% (4,407) of the documents in the three reference sources (\textit{pre-LLM} student reports, LLM-generated text, and expert writing) to the training set. The remaining 25\% (1,470) of documents in the reference sources were assigned to the test set. We stratified the train-test split by reference source so that the training and test data had approximately equal proportions of each one. We fit the overall LDA model on the training set and assessed its classification performance on the test set. Finally, we projected the \textit{since-LLM} student reports onto the resulting two-dimensional subspace. This gave us the raw scores for each document along the two linear discriminant dimensions, as well as the posterior probability that the document belongs to each of the three reference sources.

\textbf{Quintile LDA Model:} While the overall LDA model compares the writing style of each \textit{since-LLM} student report to those of the reference sources,  we hypothesized that students might consult LLMs more often when writing certain sections of the report. In particular, we posited that students may consult LLMs more often if they feel unsure about how to write the report section, or do not see how it relates to the course material. To test whether certain sections were more LLM-like, we split each student report into quintiles by length. These quintiles roughly map to the IDMRaD-style format that student reports are expected to follow (see Section \ref{subsec-student-corpus}); for example, the first quintile 
roughly maps onto the introduction section.

To fit the quintile LDA model, we first jointly standardized the Biber feature rates across fourteen text sources: the full \textit{pre-LLM} student reports, \textit{since-LLM} student reports, LLM-generated text, expert writing, and the quintiles of the \textit{pre-LLM} and \textit{since-LLM} student reports. As with the overall LDA model, we fit the model using \textit{pre-LLM} student data analysis reports, LLM-generated texts, and expert writing (though these were slightly standardized differently, since each student report was split into quintiles). We randomly assigned 75\% (4,407) of the documents from reference sources to the training set and the remaining 25\% (1,470) to the test set, stratifying by source. We fit the quintile LDA model with the training set and assessed its classification performance on the test set. Finally, we projected the \textit{pre-LLM} and \textit{since-LLM} student report quintiles onto the resulting two-dimensional subspace. This gave us the raw scores for each document along the two linear discriminant dimensions, as well as the posterior probability that the document belongs to each of the three reference sources. 

\subsubsection{Comparing Lemma Usage}
\label{subsec:lemma-usage}

In order to analyze how student lemma usage has evolved with the increasing popularity of LLMs, we first identified each LLM's ``favorite'' verbs by extracting its fifty most used verb lemmas in the LLM corpus. We chose to study verb usage over time because it was more context-independent in the student and LLM corpora, whereas nouns and adjectives more heavily depended on the paper topic. We removed all auxiliary verbs (e.g. \textit{will}, \textit{be}, \textit{can}, \textit{might}) since they are helping or modal verbs, rather than the main verbs of a sentence. 

Combining the lists of each LLM's top fifty most used verbs resulted in 97 unique lemmas, including \textit{address, develop, enhance, ensure} and \textit{influence}. We conducted a keyness analysis to study how the use of these verbs changed over time in the student corpus \citep{Gabrielatos2018}. Specifically, we tested whether each of the 97 verb lemmas occurred more often in a target corpus (the \textit{since-LLM} student reports) than a reference corpus (the \textit{pre-LLM} student reports) using a binomial likelihood-ratio test. To account for multiple testing, we determined statistical significance using the Bonferroni-adjusted p-values at a significance level of 0.05. 

\section{Results}\label{sec-results}

\subsection{Distinguishing between \textit{Pre-LLM} Student, LLM-generated, and Expert Writing Styles}
\label{subsec:distinguishing-sources}

\begin{figure}
    \centering    \includegraphics[width=0.7\linewidth]{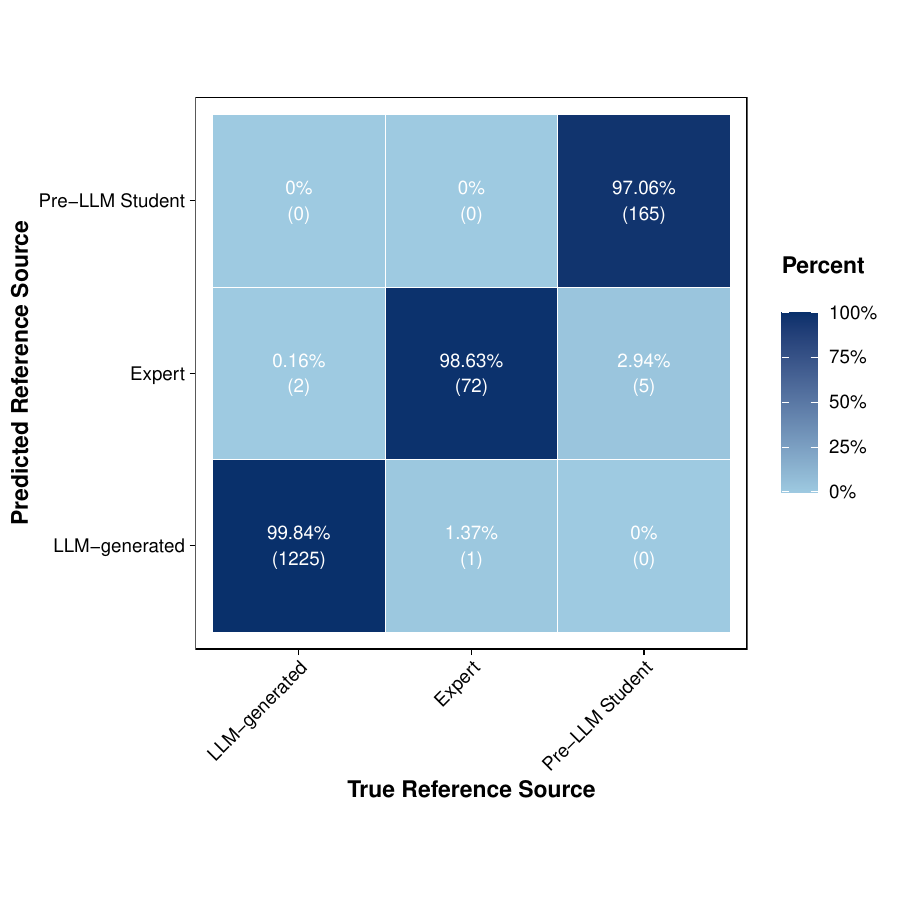}
    \caption{Confusion matrix on the test data for the overall LDA model, which compares \textit{pre-LLM} student reports to expert writing and LLM-generated text.} 
    \label{fig:confusion-matrix}
\end{figure}

There are clear stylistic differences in \textit{pre-LLM} student, LLM-generated, and expert writing styles: the overall LDA model achieves a test accuracy of 99.46\%. Per the confusion matrix in Figure \ref{fig:confusion-matrix}, none of the \textit{pre-LLM} student reports in the test set are misclassified as LLM-generated text with the overall LDA model, and only five (2.94\%) \textit{pre-LLM} student reports in the test set are misclassified as expert writing. There is also very little confusion between expert and LLM-generated text; only one (1.37\%) expert document in the test set is misclassified as LLM-generated text, and two (0.16\%) LLM-generated texts in the test set are misclassified as expert writing. 

\begin{figure}
    \centering
    \includegraphics[width=\linewidth]{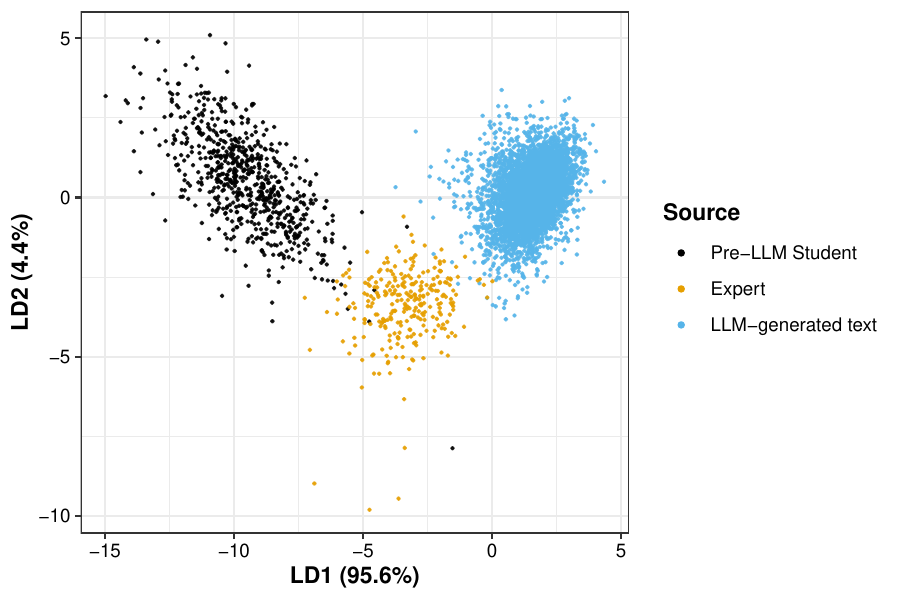}
    \caption{Projection of \textit{pre-LLM} student reports, LLM-generated text, and expert writing onto the two linear discriminants from the overall LDA model.} 
    \label{fig:LDA-ridgeplot-projection}
\end{figure}

Figure \ref{fig:LDA-ridgeplot-projection} depicts how the overall LDA model distinguishes between the three reference sources in the two-dimensional subspace. The first linear discriminant (LD1), which explains 95.6\% of the variation in the jointly standardized Biber feature rates, primarily separates LLM-generated text from \textit{pre-LLM} student reports and expert writing. Meanwhile, the second linear discriminant (LD2), which explains 4.4\% of the variation in the jointly standardized Biber feature rates, distinguishes expert writing from LLM-generated text and \textit{pre-LLM} student reports. 

\subsubsection{Writing Features Distinguishing between \textit{Pre-LLM} student, LLM-generated, and Expert Writing}
\label{subsubsec:stylistic-features}

To analyze which stylistic features most distinguish \textit{pre-LLM} student reports from LLM-generated text and expert writing, we take the union of the Biber features with the ten largest coefficients (by magnitude) in each linear discriminant of the overall LDA model. 
From Table \ref{tab:main-biber-features}, \textit{pre-LLM} student reports tend to be less informationally dense than LLM-generated text and expert writing. Several Biber features associated with \textit{pre-LLM} student reports are interactive writing features, such as first-person pronouns (e.g., \textit{I, my, we, our}), contractions (e.g., \textit{shouldn't}, \textit{could've}, \textit{can't}), and discourse particles (e.g., \textit{well, now, anyway}). These features have previously been shown to be more common in novice than expert writing \citep{Staples2016}. \textit{Pre-LLM} student reports also tend to contain higher rates of existential `there' (e.g., \textit{there} are 47 observations) and `that’ subjects (e.g., the graphs \textit{that} are presented above allow us to understand that...). Generally, existential `there' is a grammatical placeholder that is used to introduce new information or complex ideas. Meanwhile, `that' subjects can help add specificity to the noun phrase in the subject position. For instance, ``\textit{that} are presented above" adds specificity to the subject ``graphs". Both of these features tend to be less common in expert writing; instead, experts often rely more heavily on complex noun phrases to introduce new information or concepts \citep{Staples2016, Aull2017, DeLuca2025}. 

Likewise, Table \ref{tab:main-biber-features} indicates that LLM-generated text is more informationally dense than \textit{pre-LLM} student and expert writing. LLM-generated texts tend to have longer words on average, as well as higher rates of present participles (e.g., \textit{writing, analyzing, assessing}) and nominalizations (e.g., \textit{flexibility} from \textit{flexible}) than both \textit{pre-LLM} student reports and expert writing. These results align with those from \cite{Reinhart2025}, who also examine the stylistic differences between LLM-generated text and human-authored writing with Biber features. LLM-generated text also tends to have higher rates of attributive adjectives (e.g., \textit{substantive, special, possible}) and adverbs (e.g., \textit{substantively, especially, possibly}) than \textit{pre-LLM} student reports and expert writing. Often, nominalizations, attributive adjectives, adverbs, and participles produce denser noun and verb phrases, and thus more informationally dense writing \citep{Biber2011, Staples2016, Markey2024, Reinhart2025}. In turn, these results confirm that \textit{pre-LLM} student reports tend to be less informationally dense than LLM-generated writing.

Turning to expert writing, we find that experts tend to use higher average rates of present tense and agentless passives (e.g., \textit{findings are limited by the small sample size}) than \textit{pre-LLM} student reports and LLM-generated text. Expert writing also relies more on `be' as the main sentence verb (e.g., the distribution \textit{is} non-skewed) and tends to have higher rates of stranded prepositions (e.g., tools that students can engage \textit{with}) than \textit{pre-LLM} student reports and LLM-generated text. While having `be' as the main sentence verb and stranded prepositions are more common in interactive prose than academic writing \citep{Biber1988}, we explicitly targeted applied statistics articles that are more accessible to non-statistical audiences. For instance, \textit{CHANCE} magazine is explicitly intended for a ``more general audience" than statistics professionals \citep{chance-about}. Therefore, it is unsurprising that our expert corpus contains articles that are more interactive and less informationally dense than traditional academic writing. 

\begin{table}
\centering
\begin{spacing}{1.25}
\renewcommand{\arraystretch}{1.2}
\begin{tabular}{p{9cm}rr}
\toprule
Biber Feature & \makecell{LD1 Coefficient} & \makecell{LD2 Coefficient} \\
\toprule
\studentrow First-person pronouns & $-1.03$ & $0.58$ \\
\expertrow `Be' main verb & $-0.98$ & $-0.21$ \\
\studentrow Existential `there' & $-0.69$ & $0.93$ \\
\llmrow Mean word length & $0.60$ & $1.07$ \\
\studentrow `Seem' or `appear' verbs & $-0.41$ & $0.45$ \\
\llmrow Attributive adjectives & $0.25$ & $0.20$ \\
\llmrow Adverbs & $0.24$ & $0.02$ \\
\llmrow Nominalizations & $0.23$ & $0.13$ \\
\expertrow Present tense & $-0.23$ & $-0.14$ \\
\studentrow `That' subject & $-0.19$ & $0.06$ \\
\studentrow Contractions & $-0.05$ & $0.25$ \\
\llmrow Present participle & $0.13$ & $0.24$ \\
\expertrow Stranded prepositions & $-0.07$ & $-0.24$ \\
\llmrow Phrasal coordination & $0.14$ & $0.24$ \\
\studentrow Discourse particles & $-0.06$ & $0.23$ \\
\expertrow Agentless passives & $-0.10$ & $-0.22$ \\
\bottomrule
\end{tabular}
\caption{Union of the Biber features with top ten coefficient magnitudes in the first (LD1) and second (LD2) linear discriminants of the overall LDA model. Rows are shaded according to whether the feature is more associated with \textit{pre-LLM} student reports (black), LLM-generated text (blue), or expert writing (orange).}
\label{tab:main-biber-features}
\end{spacing}
\end{table}

\subsection{Trajectory of Writing Style in Students' Statistics Reports}
\label{subsec:trajectory}

\begin{figure}
    \centering
    \includegraphics[width=0.9\linewidth]{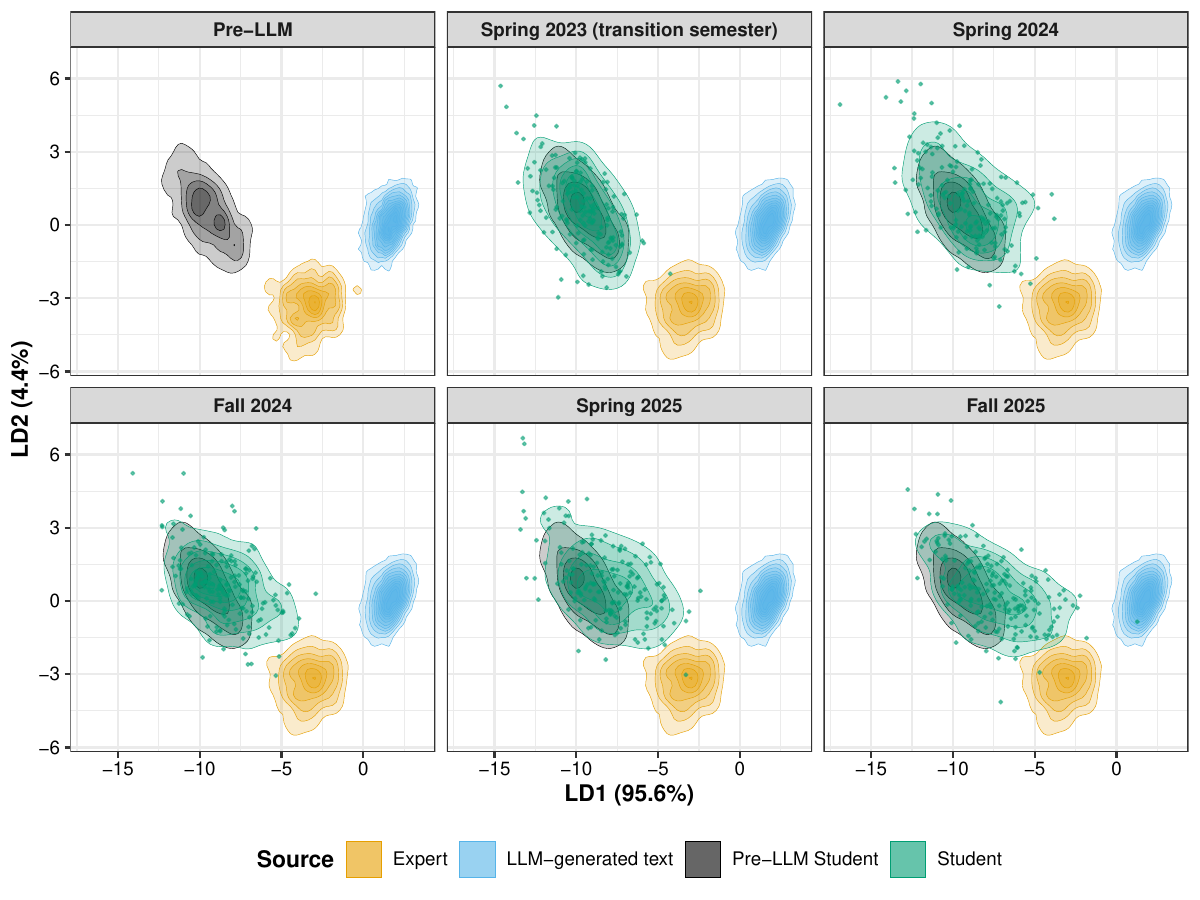}
    \caption{Contour plot of the linear discriminant scores for each text source, faceted by semester. Points are used to represent the \textit{transition year} and \textit{since-LLM} student reports. We plot the reference sources (\textit{pre-LLM} student, LLM-generated text, and expert writing) in each facet.} 
    \label{fig:LD1-LD2-contours}
\end{figure}

We illustrate how the writing style in student reports has changed since the emergence of LLMs in Figure \ref{fig:LD1-LD2-contours}. Overall, student reports have tended towards a mixture of LLM and expert writing styles since the emergence of LLMs; very few \textit{since-LLM} student reports are either fully expert-like or LLM-like.

Additionally, all quintiles of the students' reports have a lower average probability belonging to the (full) \textit{pre-LLM} student report class since LLMs became widely accessible (see Table \ref{tab:avg-post-prob-preLLM}). It is worth noting, however, that some sections of student reports were more LLM-like or expert-like than others, even before ChatGPT was released in 2022. In particular, the first quintile of \textit{pre-LLM} student reports had a 32.6\% average posterior probability of belonging to the (full) \textit{pre-LLM} student reports class. This implies that the writing style at the beginning of \textit{pre-LLM} student reports tends to differ from the typical writing style in the rest of \textit{pre-LLM} students' reports. 

\begin{table}
\centering

\rowcolors{3}{}{lightgray}
\begin{tabular}{lrrr}
\toprule
& \multicolumn{2}{c}{Mean posterior probability (SD)} & \\
\cmidrule(lr){2-3}
Report quintile & \textit{Pre-LLM} & \textit{Since-LLM} & Percent change \\
\midrule
First (introduction) & 32.6 (44.4) & 26.9 (42.3) & $-17.5$ \\
Second (data)        & 91.5 (26.9) & 85.6 (33.8) & $-6.4$ \\
Third (methods)      & 97.2 (15.4) & 93.5 (23.9) & $-3.8$ \\
Fourth (results)     & 94.3 (22.3) & 87.5 (31.8) & $-7.2$ \\
Fifth (discussion)   & 75.2 (40.9) & 61.0 (47.2) & $-18.8$ \\
\bottomrule
\end{tabular}
    \caption{Average (SD) posterior probability of belonging to the reference class of full pre-LLM student statistics' reports for each report quintile, \textit{pre-LLM} versus \textit{since-LLM}. The percent change is the difference in the mean posterior probability \textit{pre-LLM} versus \textit{since-LLM}, divided by the mean posterior probability \textit{pre-LLM}. The parenthetical after the report quintile indicates the associated IDMRaD section.} 
    \label{tab:avg-post-prob-preLLM}
\end{table}

\begin{figure}
    \centering
    \includegraphics[width=0.9\linewidth]{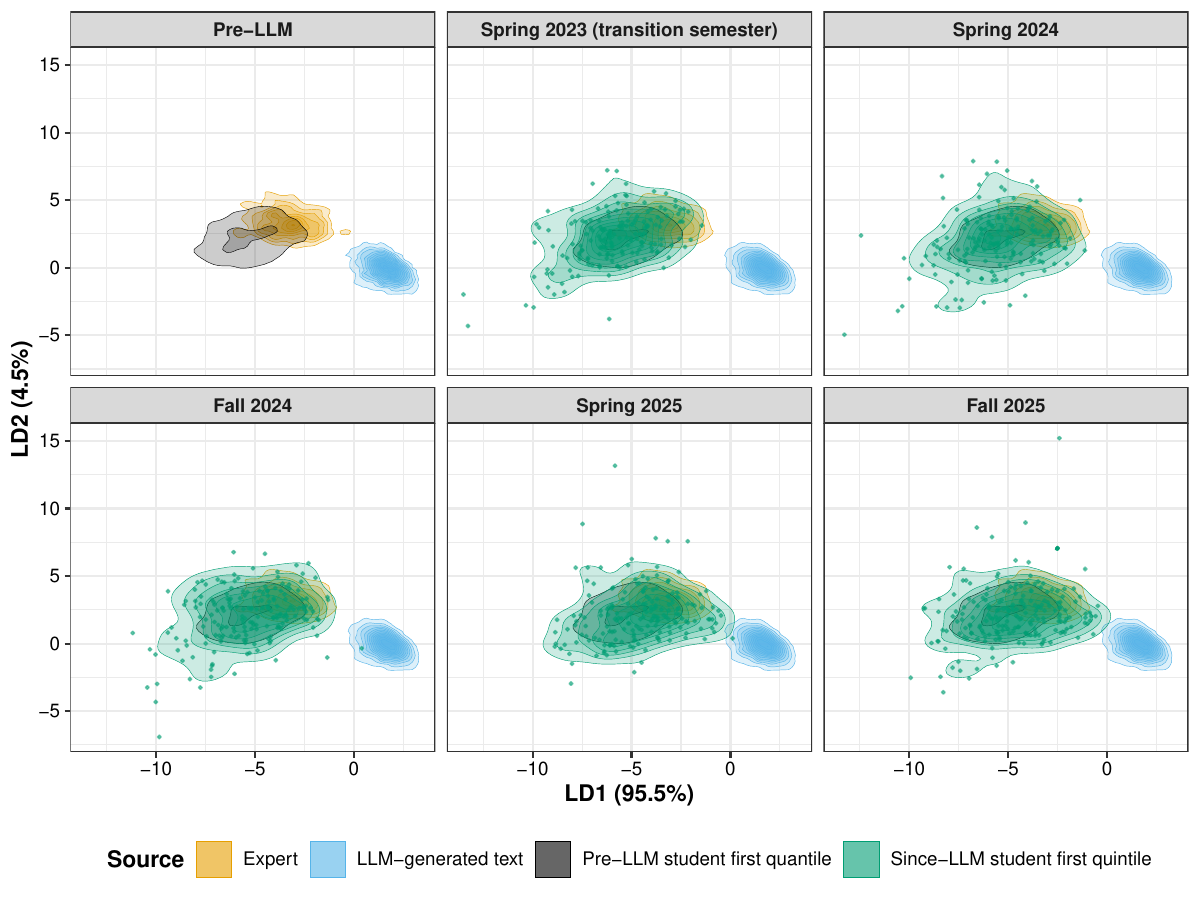}
    \caption{Contour plot of the first and second linear discriminant scores for each text source, faceted by academic semester. Each point represents the fifth quintile of a student report, which roughly maps onto the discussion section. We plot the reference sources (\textit{pre-LLM} student, LLM-generated text, and expert writing) in each facet.} 
    \label{fig:LD1-LD5-contours-introduction}
\end{figure}

The largest percent change in the average posterior probability of belonging to the (full) \textit{pre-LLM} student report class occurs in the first and last quintiles, which correspond to the introduction and discussion sections, respectively. Using the quintile LDA model, we visualize the writing style trajectory of the first quintile in Figure \ref{fig:LD1-LD5-contours-introduction}. Prior to LLMs, the writing style in students' first quintiles overlaps with that of experts. The average writing style in students' first quintile shifts after the emergence of LLMs: simultaneously becoming more expert-like and LLM-like. By Spring 2025, the writing style in the first quintile of some students' statistics reports has begun to overlap that of LLM-generated text.   

\begin{figure}
    \centering
    \includegraphics[width=0.9\linewidth]{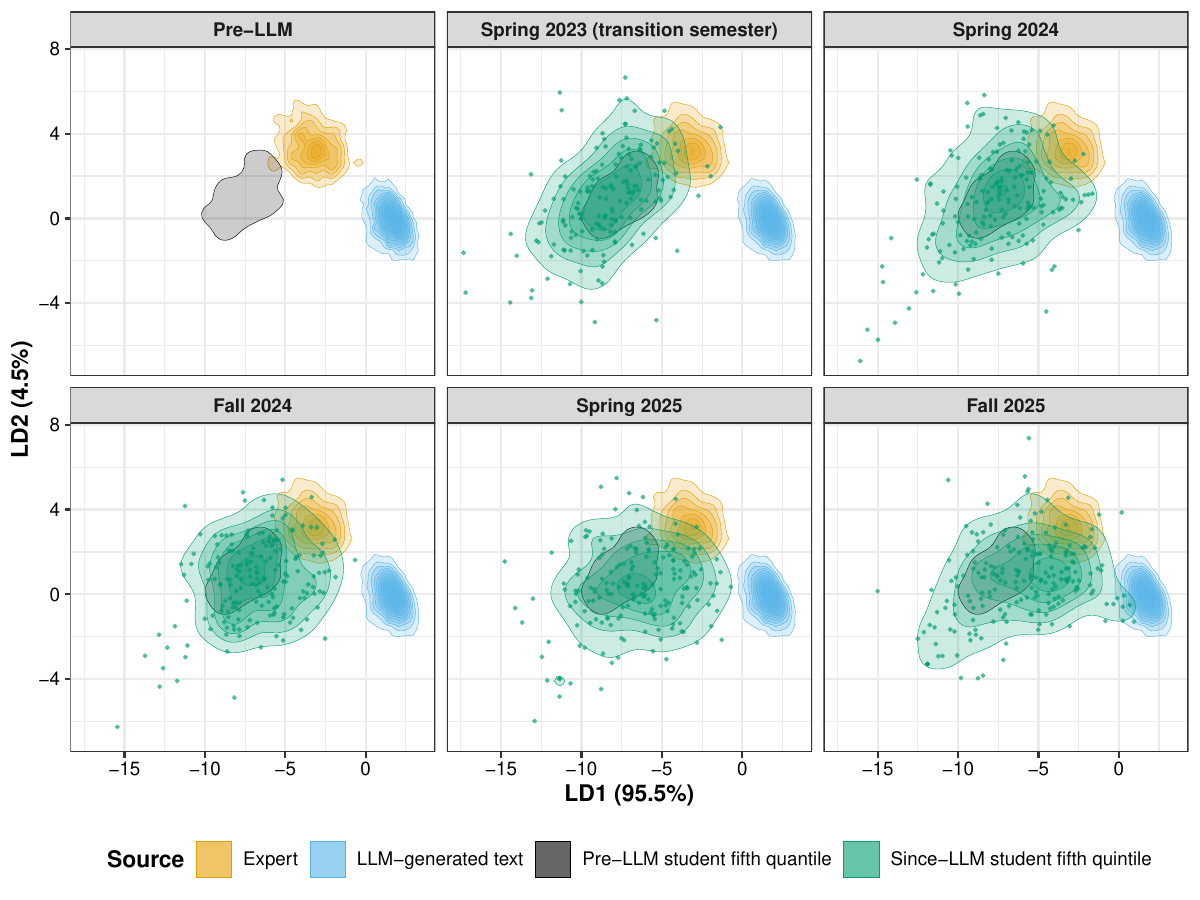}
    \caption{Contour plot of the first and second linear discriminant scores for each text source, faceted by academic semester. Each point represents the first quintile of a student report, which roughly maps onto the introduction section, and the color denotes the text source. We plot the reference sources (\textit{pre-LLM} student reports, LLM-generated text, and expert writing) in each facet.} 
    \label{fig:LD1-LD5-contours-conclusion}
\end{figure}

Figure \ref{fig:LD1-LD5-contours-conclusion} shows the trajectory of the writing style in students' fifth quintiles. Similar to the first quintile, there is some overlap between the writing styles of students' fifth quintile and expert writing before LLMs. In the \textit{since-LLM} semesters, the writing style in the fifth quintile of students' reports has tended towards a mixture of expert and LLM writing styles. By Fall 2025, some students' fifth quintiles are fully expert-like or fully LLM-like, though most appear to be a mixture of the two styles.

\subsubsection{Shift in Prominent Writing Features since the Emergence of LLMs}

Consistent with the trajectory results in Section \ref{subsec:trajectory}, many stylistic features associated with \textit{pre-LLM} student reports are used less frequently in student reports after the emergence of LLMs. In particular, there are lower rates of first-person pronouns, the verbs `seem' and `appear', existential `there' clauses, and `that' subject clauses in \textit{since-LLM} than \textit{pre-LLM} student reports (see  Table \ref{tab:largest-diff}). 

However, instead of adopting the stylistic features associated with expert writing, students seem to be mostly adopting those of LLM-generated text. All six of the sixteen Biber features from Table \ref{tab:main-biber-features} that are associated with LLM-generated text occur more frequently in \textit{since-LLM} student reports. Moreover, of the sixteen Biber features, the top four with the largest standardized mean shifts \textit{pre-LLM} to \textit{since-LLM} are associated with LLM-generated text. In contrast, the stylistic features that are most indicative of expert writing, such as `be' as the main verb and agentless passives, have lower usage rates in \textit{since-LLM} student reports. This implies that students tend to adopt the stylistic features associated with LLM-generated text directly, resulting in more informationally dense reports. On the other hand, students tend not to directly adopt the prominent stylistic features associated with expert writing and thus may only be producing more expert-like writing incidentally by picking up on LLMs' stylistic features.

\begin{table}
\centering
\begin{spacing}{1.25}
\renewcommand{\arraystretch}{1.2}
\begin{tabular}{p{8.5cm}r}
\toprule
Biber feature & \makecell{Standardized mean shift \\in \textit{since-LLM} student reports}\\
\toprule
\llmrow Present participles & 0.93 \\
\llmrow Mean word length & 0.89 \\
\llmrow Attributive adjectives& 0.50	\\
\llmrow Nominalizations & 0.41 \\
\expertrow `Be' main verb & $-0.31$ \\
\studentrow First-person pronouns & $-0.28$ \\
\studentrow Existential `there' & $-0.27$ \\
\expertrow Agentless passives & $-0.23$ \\
\studentrow Verbs `seem' or `appear' & $-0.12$ \\
\studentrow Discourse particles & $0.10$ \\
\studentrow `That' subject & $-0.07$ \\
\expertrow Stranded preposition & $-0.07$ \\
\llmrow Adverbs & $0.05$ \\
\expertrow Present tense & $-0.04$ \\
\llmrow Phrasal coordination & $0.04$ \\
\studentrow Contractions & $0.01$ \\
\bottomrule
\end{tabular}
\end{spacing}
\caption{The union of the ten Biber features with the largest contribution to each linear discriminant of the overall LDA model, and the corresponding standardized mean shift of the usage rate in \textit{pre-LLM} versus \textit{since-LLM} student reports. The standardized mean shift is calculated as the difference in mean rate for each feature in the \textit{since-LLM} and \textit{pre-LLM} student reports, divided by the standard deviation of the feature in the \textit{pre-LLM} student reports. Rows are shaded according to whether the feature is more associated with \textit{pre-LLM} student reports (black), LLM-generated text (blue), or expert writing (orange).}
\label{tab:largest-diff}
\end{table}

\subsection{Changes in the Frequency of LLMs' ``Favorite'' Verbs in Students' Statistics Reports}

\begin{figure}
    \centering
    \includegraphics[width=\linewidth]{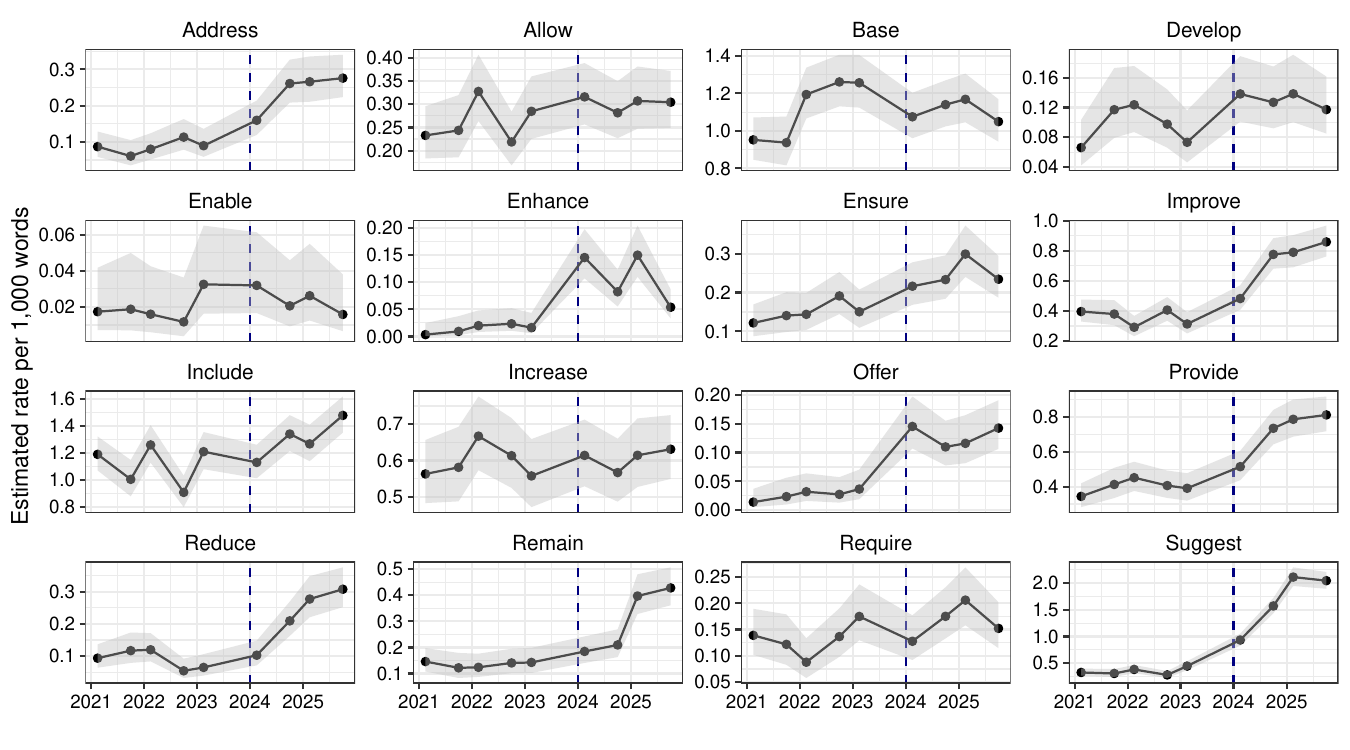}
    \caption{The rate of use in the student corpus of sixteen lemmas that appeared in the top fifty most used verbs for each LLM, aside from \textit{associate}. The vertical line represents the beginning of the \textit{since-LLM} period.} 
    \label{fig:llm_verbs}
\end{figure}

Seventeen lemmas appeared in the top fifty most frequently used verbs for all four LLMs (GPT-4o, GPT-5 Mini, Gemini Flash 2.5, and Claude Haiku 4.5). Figure \ref{fig:llm_verbs} shows the trajectory of the usage rates of sixteen of these verbs in students' reports. We omit \textit{associate}, which showed no statistically significant change in usage, due to its statistical connotation. All sixteen verbs increased in usage in the student corpus during the \textit{since-LLM} era relative to the \textit{pre-LLM} era, with ten showing statistically significant increases according to our keyness analysis with Bonferroni-corrected $p$-values. Looking at the 97 verbs as a whole, 39 (40.2\%) exhibited statistically significant positive change, 4 (4.1\%) showed statistically significant negative change, and 53 (54.6\%) displayed no statistically significant change in usage rate when comparing \textit{since-LLM} reports to \textit{pre-LLM} reports. Particularly striking are \textit{enhance}, which increased from only fourteen occurrences in \textit{pre-LLM} student reports to 122 in \textit{since-LLM} reports (an 8.71-fold increase), and \textit{suggest}, which more than quintupled in rate of use in \textit{since-LLM} reports relative to \textit{pre-LLM} reports.

The verb \textit{underscore} is also particularly interesting. A keyness analysis could not be conducted on this lemma because it was never used in \textit{pre-LLM} student reports (the reference corpus); however, it was used 35 times in \textit{since-LLM} student reports. A concordance analysis for \textit{underscore}, which extracts the four words preceding and following the lemma of interest, is shown in Figure~\ref{fig:concordance}. Phrases such as ``\textit{underscored} by an F-test" and repeated usage patterns like, ``\textit{underscores} the complexity of" give us strong evidence that the recent rise in \textit{underscore} is largely driven by LLM use. We find similarly interesting patterns with a concordance analysis on \textit{offer} and \textit{provide}. In \textit{pre-LLM} reports, \textit{offer} was used primarily in specific contextual phrases, such as ``\textit{offering} their [bike rental] services". In contrast, in \textit{since-LLM} reports, students more frequently use \textit{offer} and \textit{provide} in more promotional language, such as ``...\textit{offer} valuable perspectives", ``numerical summaries \textit{offer}...", and especially, ``\textit{provide} valuable insights into...".

\begin{figure}
\centering
\renewcommand{\arraystretch}{1}
\begin{tabular}{r@{\hspace{0.2em}}c@{\hspace{0.3em}}l}
 met. Overall, this analysis &\textbf{underscores}& the complexity of predicting \\
 predictors. Overall, this analysis &\textbf{underscores}& the value of data \\
 of this model is &\textbf{underscored}& by an F-test \\
 reference day. This finding &\textbf{underscores}& the critical role that \\
 model's significance is &\textbf{underscored}& by the regression F \\
 as a significant factor, &\textbf{underscored}& by a p-value \\
 affect sharing counts. This &\textbf{underscores}& the complexity of social \\
 media engagement. Our study &\textbf{underscores}& the complexity of social \\
 exorbitant cost of living &\textbf{underscores}& the critical need to \\
 effectively. Ultimately, this analysis &\textbf{underscores}& the complex interplay of \\
 analysis of error diagnostics &\textbf{underscored}& that our model meets \\
 results we wanted, it &\textbf{underscores}& the importance of considering \\
 practical standpoint, the prediction &\textbf{underscores}& that articles meeting these \\
  modest predictive power, it &\textbf{underscores}& an important relationship between \\
  growth of shared mobility &\textbf{underscores}& the ongoing importance of
\end{tabular}
\caption{Fifteen ways in which \textit{underscore} was used in \textit{since-LLM} student reports. The lemma \textit{underscore} was never used in \textit{pre-LLM} student reports.}
\label{fig:concordance}
\end{figure}

\section{Conclusion}\label{sec-conc}

Through our analysis, we identified stylistic differences between  \textit{pre-LLM} student statistics reports, expert applied statistics writing, and LLM-generated text. In particular, \textit{pre-LLM} student reports tend to be more interactive and less informationally dense than expert writing and LLM-generated text, which aligns with previous research findings \citep{Staples2016, Reinhart2025, DeLuca2025}. With the rise of LLMs, we find that the writing style in students' statistics reports has shifted toward a \textit{mixture} of LLM and expert writing styles. This shift is particularly pronounced in the first and fifth quintiles of students' reports, which correspond to the introduction and conclusion sections, respectively. Generally, \textit{since-LLM} student reports tend to have longer words and higher rates of present participle clauses, attributive adjectives, and nominalizations than \textit{pre-LLM} student reports. In line with \cite{Reinhart2025}, these features are indicative of LLM-generated text. Finally, many of the LLMs' favorite verbs, such as \textit{suggest} and \textit{offer}, are used more frequently in \textit{since-LLM} student reports than \textit{pre-LLM} student reports. Taken together, these results provide strong evidence that students' statistics reports have become more LLM-like in terms of both style and verb usage. At the same time, we find evidence that students' reports have become more expert-like in terms of writing style since the introduction of LLMs.

The most likely explanation for our results is that students are consulting LLMs while writing their statistics reports. While it could be that students are mimicking AI-generated content they see in online media, social platforms, or academic writing (e.g., see \citealp{llmscience, Sun2024, Ansari2025}), many undergraduate students report using LLMs in their assignments. For instance, a survey from March to August 2024 of over 95,000 students from twenty major public research-intensive universities in the United States found that two-thirds of students reported using generative AI during the academic year, with 37\% using it regularly \citep{Chirikov2026}. Hence, we expect that at least some of our observed shifts are attributable to students consulting LLMs while writing their reports.

Given this, we suspect that the large stylistic shift in the first and fifth quintiles of students' reports is a direct result of students consulting LLMs to help write their introductions and conclusions. This could be because students (i) feel unsure about how to write (or structure) these sections, or (ii) view writing the middle report sections (i.e., the exploratory data analysis, methods, and results sections) as more statistically valuable than writing the introduction and conclusion sections. In Section \ref{subsec:implications}, we suggest assessments that explicitly target the introduction and conclusion sections of students' reports. 

Finally, while evidence that students' statistics writing has become more expert-like in recent semesters might be exciting, its value depends on the function that writing assignments are supposed to serve in the statistics and data science curriculum. As emphasized by \cite{DeLuca2025}, the primary goal of writing-to-learn assignments is not that students produce polished, expert-like reports, but that writing helps expose and resolve their conceptual misunderstandings. In this context, improvements in writing style provide little educational value if they stem from students consulting LLMs and bypassing key aspects of the learning process. On the other hand, it might seem like producing more expert-like writing is beneficial for writing-in-the-disciplines assignments, where students tailor the communication of their statistical results to less-statistical audiences, such as stakeholders and domain experts. However,  we do not account for the \textit{statistical quality} of students' reports in our analysis. Thus, it is possible that the writing style in students' reports has become more similar to that of experts, but the statistical reasoning in the reports has diminished with the emergence of LLMs.

\subsection{Implications for Statistics and Data Science Educators}
\label{subsec:implications}

The first recommendation of the GAISE 2016 College Guidelines is to ``teach statistical thinking" \citep{gaise}. To that end, statistics and data science curricula should provide students with opportunities to think critically about statistical issues and emphasize the practical problem-solving skills necessary to answer statistical questions. In practice, these problem-solving skills require students to make connections between statistical concepts and recognize that statistical questions can be answered using a variety of methods or procedures \citep{gaise}. While statistical thinking skills are, of course, useful for students who plan to pursue graduate school or careers in statistics and data science, we believe that they are crucial for \textit{all} students. Regardless of their career path, students will be constantly exposed to data and statistical results in their daily lives \citep{Rumsey2002}, for instance, in sports betting, political polling, or disease forecasting contexts, among many others. Developing strong statistical thinking skills enables students to be critical consumers of the data and statistical claims that will inevitably be presented to them.

The recommendation to ``teach statistical thinking" is more important than ever in the age of LLMs and other generative AI tools. Even if LLMs can assist with coding or writing tasks, students must still be able to interpret results, evaluate assumptions, assess quality, and evaluate the statistical, practical, and ethical implications of the LLM's output. Reflecting on statistics education in the age of AI, ASA President Jeri Mulrow emphasized that adapting curricula to new technologies ``does not mean abandoning the fundamentals" \citep{Mulrow2026}. Rather, it heightens the importance of developing students' critical judgment and statistical thinking skills needed to use these tools responsibly. 

Yet, the way in which statistical thinking is taught and assessed requires careful reevaluation in the age of AI. Using traditional project-based assessments, such as take-home writing assignments, to assess students' statistical thinking is fallible given the accessibility of generative AI; it is no longer clear how much of a student's work is their own versus from an LLM. As a result, some statistics and data science instructors have decided to replace all project-based assessments with in-class exams. However, we do not believe that the statistical thinking skills---such as end-to-end data analysis and communication skills---that project-based learning helps develop can be fully replicated by in-class exams. Thus, the challenge becomes how to retain the important aspects of project-based assessments while also acknowledging that students may consult LLMs.

\subsubsection{Alternative Modes of Assessment}

A potential alternative to in-class exams is oral exams. Two benefits of oral exams are that they help develop communication skills and provide a more authentic assessment experience, since students are unlikely to complete written exams after graduation \citep{Theobold2021}. Additionally, students explain and apply concepts in their own words during an oral exam, which allows the instructor to assess a student's statistical thinking without AI assistance. Still, there are limitations to oral examinations. They often require a substantial time commitment from instructors and teaching assistants, particularly in larger classes. Oral exams may also be a source of anxiety for students not familiar with this form of assessment \citep{Theobold2021}.

Another suggestion is to incorporate smaller, targeted writing assignments or short quizzes, where students must devise an end-to-end data analysis plan or write a small section of a data analysis report (e.g., a description of a dataset or plot, interpretation of a coefficient, or the limitations of an analysis) following a certain structure. For example, \cite{Woodard2020} proposed a four-step structure in which students answer the question, state the relevant facts from the problem, state the implications of those facts, and explain how those facts lead to the conclusion. More broadly, explicit rhetorical instruction and pattern practice exercises have been shown to improve students' scientific writing and strengthen their awareness of audience and genre expectations \citep{Wolfe2011}. To target the introduction section of data analysis reports, for instance, instructors could provide students with a dataset, background information, and statistical results, and ask them to write an introduction in which they explain the significance of the research problem, describe the current state of knowledge, identify a gap, and finally position the present work as a response to that gap \citep{swales}. This assessment helps students better recognize the importance of the introduction section and how it should be structured. Hence, they may be less tempted to offload introduction writing tasks to an LLM in future work.

Our final suggestion could be either an assessment or an in-class exercise that probes students' statistical thinking as they use LLMs. Developing the ability to assess AI-generated statistical reasoning aligns with calls to make statistical thinking and evaluation a core component of digital literacy in schools \citep{rssAI}. Thus, it would be worthwhile to guide students through critiquing LLM-generated output for unstated assumptions or potential sources of error \citep{Ellis2023, Xing2024}. In doing so, students will be encouraged to evaluate the appropriateness of the LLM's chosen statistical methodology, recall the assumptions underlying that method, and discuss the statistical, ethical, and practical implications and limitations presented by the LLM. We expect this exercise to be particularly valuable, as LLMs often fail to address all critical elements of statistical tasks \citep{Xing2024}, thereby highlighting the shortcomings of overreliance on these tools.

\subsection{Limitations and Future Work}

While we find strong evidence that students' statistics writing has shifted in both style and verb usage since the emergence of LLMs, our results are based on a sample of student reports from a single introductory course at \versionphrase{}{}. The extent to which students use generative AI for writing may depend on several factors, including the type of institution, class size, and the course's policy on generative AI usage. Additionally, our LLM-generated corpus may not reflect the output that students would receive when using LLMs. Students may choose different prompts, perhaps pulling language specifically from the assignment directions or other class materials. Realistically, we would also expect students to engage conversationally with the LLM. For instance, a student might ask for specific updates to the LLM's output or feed the LLM a drafted paragraph and ask it to clean up the grammar or fix the writing style so that it sounds more professional. Therefore, it is unlikely that our LLM corpus captures the exact writing style that a student prompting the tool will receive. 

Another limitation is that we do not measure the statistical quality of students' reports in our analysis. In an effort to better understand how a more expert-like writing style corresponds to statistical understanding, future work could incorporate student grades or the reports' statistical quality into our analysis. Another interesting direction for future work would be to examine students' writing styles longitudinally (either within a course or across several courses). Through a longitudinal analysis, researchers could evaluate how students' ability to communicate statistical concepts evolves with access to generative AI, and how this differs from the evolution of students' written communication skills before LLMs.

\ifnum\anon=1
\section*{Acknowledgments}

Thank you to members of the Statistical Pedagogy \& Educational Research TeachStat Group at Carnegie Mellon University for constructive feedback on this project. We are also grateful for the questions and feedback we received while presenting an earlier version of this work at the United Statistics Conference on Teaching Statistics 2025, Research Satellite.
\fi

\section*{Disclosure statement}\label{disclosure-statement}

The authors report there are no competing interests to declare.

\section*{Data Availability Statement}\label{data-availability-statement}

The expert and student data, along with the code used for corpus preprocessing and figure generation, that support this study's findings are openly available on the Open Science Framework (OSF) at \url{https://osf.io/2r8ym/overview?view_only=3f018adb2ff147139d74527505252af0}. For copyright and privacy reasons, the expert and student texts are only deposited in aggregated form in the database, but the full LLM-generated text used in this study is openly available in Hugging Face at \href{https://huggingface.co/datasets/browndw/human-ai-parallel-corpus}{10.57967/hf/3770}.

\bibliography{references}

\phantomsection\label{supplementary-material}
\bigskip

\newpage 

\begin{center}
{\large\bf SUPPLEMENTARY MATERIAL}
\end{center}

\appendix
\renewcommand{\thefigure}{\Alph{section}.\arabic{figure}}
\setcounter{figure}{0} 

\section{Biber Feature Definitions and Examples}

\begin{spacing}{1}
{\renewcommand{\arraystretch}{1.2}
\begin{longtable}{|>{\raggedright\arraybackslash}p{5cm}| >{\raggedright\arraybackslash}p{5cm}| >{\raggedright\arraybackslash}p{6.5cm}|}
\caption{Descriptions of the 67 Biber features, with examples from the student corpus.}
\label{tab:biber-features}
\\
\hline
Biber Feature & Description & Example \\
\hline
\endfirsthead

\caption[]{\textit{continued}}\\
\hline
Biber Feature & Description & Example \\
\hline
\endhead

\hline
\endfoot

\bottomrule
\endlastfoot
Past tense & Verbs in past tense & The Q–Q plot \textit{indicated} approximate normality of errors. \\
\hline
Perfect aspect & Verbs in perfect aspect & We \textit{have} uncovered strong skew in several of our variables. \\
\hline
Present tense & Verbs in present tense & We \textit{examine} the relationship between total household income and three explanatory variables. \\
\hline
Place adverbials & Adverbs and adverbial phrases describing place & here, there, above, beside, outside \\
\hline
Time adverbials &  Adverbs and adverbial phrases describing time & Recently, always, often, sometimes, never \\
\hline
First-person pronouns & & I, my, we, our \\
\hline
Second-person pronouns & & you, your \\
\hline
Third-person pronouns & (excluding \textit{it}) & he, his, she, her, they, theirs \\
\hline
Pronoun `it' & & it, its, itself \\
\hline
Demonstrative pronouns & Pronouns replacing nouns & \textit{This} suggests that the independence is justified. \\
\hline
Indefinite pronouns & & Everyone, somebody, nothing \\
\hline
Pro-verb `do' & &  We will first check for multicollinearity. To \textit{do} this, we will check the pairs plot for strong correlations. \\
\hline
`Wh-' questions & Direct who, what, when, where, and why questions & \textit{What} factors best predict household income in New York City?\\
\hline
Nominalizations & Nouns ending in \textit{-tion}, \textit{-ness}, \textit{-ment}, etc. & The \textit{distribution} for the variable `Temp' appears to be roughly bimodal. \\
\hline
Gerunds & Participial forms functioning as nouns & This paper will address the above to provide a better \textit{understanding} of the housing landscape in the city. \\
\hline
Other nouns & All other nouns & model, variable, relationship, plot, distribution \\
\hline
Agentless passives & Agentless passive voice & The predicted income \textit{is computed} as follows... \\
\hline
`By-' passives & Passive voice with agent & In a linear model where shares \textit{is explained only by content}, the $R^2$ value is approximately 0.002. \\
\hline
`Be' as main verb & Use of \textit{be} forms as main verb & The distribution \textit{is} unimodal and skewed right. \\
\hline
Existential `there' & \textit{There} used to assert something exists &  In the data given, we see that \textit{there} exist 4 variables and 299 observations. \\
\hline
`That' verb complements & \textit{that} relative clauses, which modify a predicative adjective & We see \textit{that the data is unimodal from 0-5 and has no outliers}. \\
\hline
`That' adjective complements &  \textit{that} relative clauses, which serve as the direct object complement of a verb & The beta values for content and images were so small \textit{that they did not have much effect on the predictions of our model}. \\
\hline
`Wh-' clauses & Clauses beginning with \textit{wh-} words (e.g., who, what, when) & Our goal is to understand \textit{what factors might predict a household’s income}. \\
\hline
Infinitives & Uninflected verb preceded by \textit{to} &  All three independent variables seem \textit{to show} weak correlations with income. \\
\hline
Present participial clauses & Adverbial clauses used as present participles & \textit{After visualizing our data through histograms}, we're next going to add in numerical summaries... \\
\hline
Past participial clauses & Adverbial clauses used as past participles &  \textit{Given the $R^2$ value of 0.05351}, the model is very weak. \\
\hline
Past participial postnominal & Reduced relative past participial clauses & The distribution of total damages \textit{requested by plaintiffs}... \\
\hline
Present participial postnominal & Reduced relative present participial clauses & A model \textit{utilizing just one predictor variable}... \\
\hline
`That' clauses as subject & \textit{That} relative clauses in subject position & The graphs \textit{that} are presented above allow us to understand that...  \\
\hline
`That' clauses as object & \textit{That} relative clauses in object position & we observed the number of shares \textit{that} an article gets is related to which day of the week it is published \\
\hline
`Wh-' relatives as subject & \textit{Wh-} relatives in subject position &  Such analyses will continue to be important for those \textit{who seek civil suits}. \\
\hline
`Wh-' relatives as object & \textit{Wh-} relatives in object position & The data \textit{which we analyze in this paper}... \\
\hline
Pied-piping relative clauses & Relative clauses moved in sentence by \textit{wh-} questions &  The years \textit{in which respondents moved to New York City range from 1942 to 2004}. \\
\hline
Sentence relatives & Relative clause that refers to the entire preceding clause or proposition &  None of our variables have a VIF of greater than 2.5, \textit{which means that we do not need to be concerned about multicollinearity}. \\
\hline
Because & Causative adverbial subordinator (\textit{because})& Next, \textit{because} we are interested in creating a linear regression...\\
\hline
Though & Concessive adverbial subordinators (\textit{although}, \textit{though}) & Even \textit{though} the new scatterplot does not show any distinguishable difference...\\
\hline
If, unless & Conditional adverbial subordinators (\textit{if, unless}) & Therefore, it has heteroskedasticity, and a linear model may not be preferred \textit{unless} doing some data transformations \\
\hline
Other adverbial subordinators & Other adverbial subordinators (\textit{since, while, whereas)}  & However, \textit{since} the tails are not very curvy...\\
\hline
Prepositional phrases & Group of words that begins with a preposition and ends with a noun, pronoun, or clause& \textit{In the model,..}. There are no missing values \textit{in the dataset}. We calculate the p-value \textit{under the null hypothesis}. \\
\hline
Attributive adjectives & Adjective that is directly adjacent to the noun or pronoun it modifies & Thus, future studies can explore more \textit{diverse} variables...\\
\hline
Predicative adjectives & Adjectives that follow a linking verb and describe the subject & We conclude that the model is \textit{significant}. \\
\hline
Adverbs & Words that modify verbs, adjectives, other adverbs, or sentences & On the residual plot, residuals are \textit{patternlessly} scattered above and below the zero line without any obvious pattern.\\
\hline
Type-token ratio & Ratio of unique words (types) to total number of words (tokens) & \\
\hline
Mean word length & Mean number of letters in each word & \\
\hline
Conjuncts & Words, phrases, or clauses linked together in a sentence & \textit{However}, as the correlation coefficient of income and age was very low ...\\
\hline
Downtoners & Modifiers that tone down a verb & Some of the variables \textit{barely} have any relation with Income at all. \\
\hline
Hedges & Markers of probability or uncertainty & Thus, it has seemingly \textit{almost} no correlation...\\
\hline
Amplifiers & Modifiers increasing the force of a verb & For example, our $R^2$ for this model is extremely low... \\
\hline
Emphatics & Modifiers indicating the presence of certainty & The distribution of the square root of the number of shares is less skewed and more symmetric than \textit{just} the number of shares.\\
\hline
Discourse particles & Words and phrases used to manage the flow, structure, and emotional tone of a conversation & \textit{Now} that we have identified the trends of the individual variables...\\
\hline
Demonstratives & Words used to point out specific nouns or pronouns & Through \textit{this} data, bike sharing services are...\\
\hline
Possibility modals & Modal verbs describing possibility & However, we still saw issues in the outliers which \textit{might} be worth investigating. \\
\hline
Necessity modals & Modal verbs describing necessity & We \textit{should} now check the QQ plot. \\
\hline
Predictive modals & Modal verbs describing prediction & There is a slight curvature which we \textit{will} further analyze in the next step. \\
\hline
Public verbs & Verbs describing actions that can be observed publicly, such as speaking or announcing & We \textit{remark} that the two quantitative variables, content and images, were excluded from the model.\\
\hline
Private verbs & Verbs describing private intellectual states or intellectual acts& Our analysis of court cases has allowed us to \textit{realize} that the total amount of damages...\\
\hline
Suasive verbs & Verbs describing intentions to cause a future change & We then computed the VIFs among our quantitative variables to \textit{ensure} that there is no multicollinearity. \\
\hline
`Seem' and `appear' & Verbs \textit{seem} and \textit{appear} & There \textit{appears} to be a weak positive correlation between...\\
\hline
Contractions & & \textit{can't, won't, isn't, don't, didn't,} etc.\\
\hline
`That' deletion & Subordinator \textit{that} deletion & This plot may suggest fewer people casually use bikeshare services... \\
\hline
Stranded prepositions & Prepositions stranded at the end of a sentence or clause & The model that the conclusion was based on performed \textit{well}. \\
\hline
Split infinitives & An adverb or phrase is placed between the word ``to" and the base verb of an infinitive & We aimed \textit{to more accurately estimate} the effect size. \\
\hline
Split auxiliaries & An adverb or adverbial phrase is placed between an auxiliary verb and the main verb & Now that we \textit{have already explored} the dataset...\\
\hline
Phrasal coordination & Phrases joined with coordinating conjunctions  & The next step is to look at the residual plot \textit{and} the Q-Q plot. \\
\hline
Clausal coordination & Independent clause coordination with \textit{and} & Despite this, the data we used seems to be the best option \textit{and} we can look past this violation.\\
\hline
Synthetic negation & Negation that is expressed within a single word through the addition of a negative affix, rather than through a separate word like \textit{not}. &  The response rate is \textit{unknown} and the issue of response bias lurks. \\
\hline
Analytic negation & Negation expressed through a separate negative word, rather than by attaching a negative affix to another word. &  It seems like there might be a slightly negative trend, but it's \textit{not} significant. \\
\end{longtable}}
\end{spacing}

\includepdf[pages={1-8}]{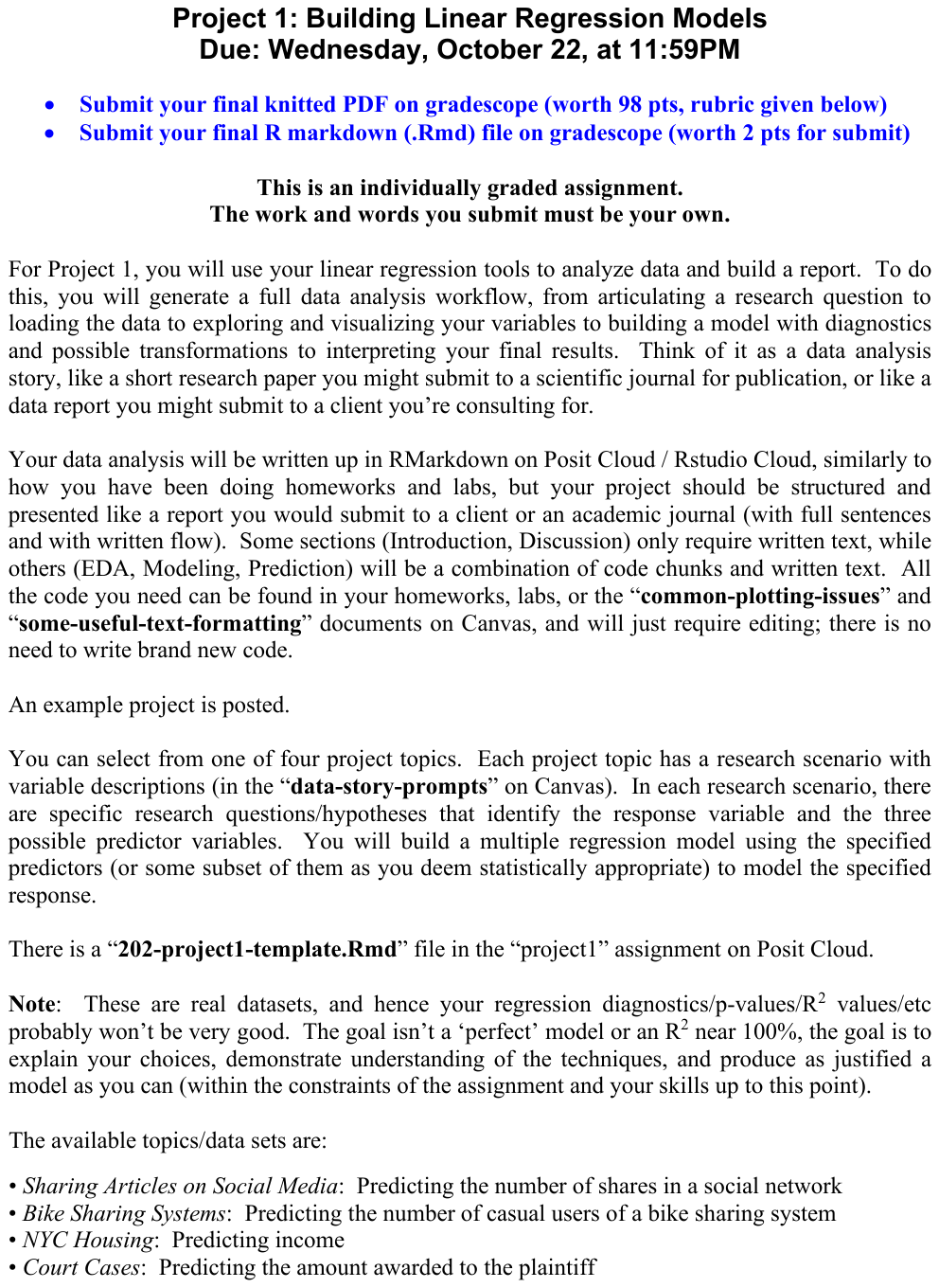}

\includepdf[pages = 1]{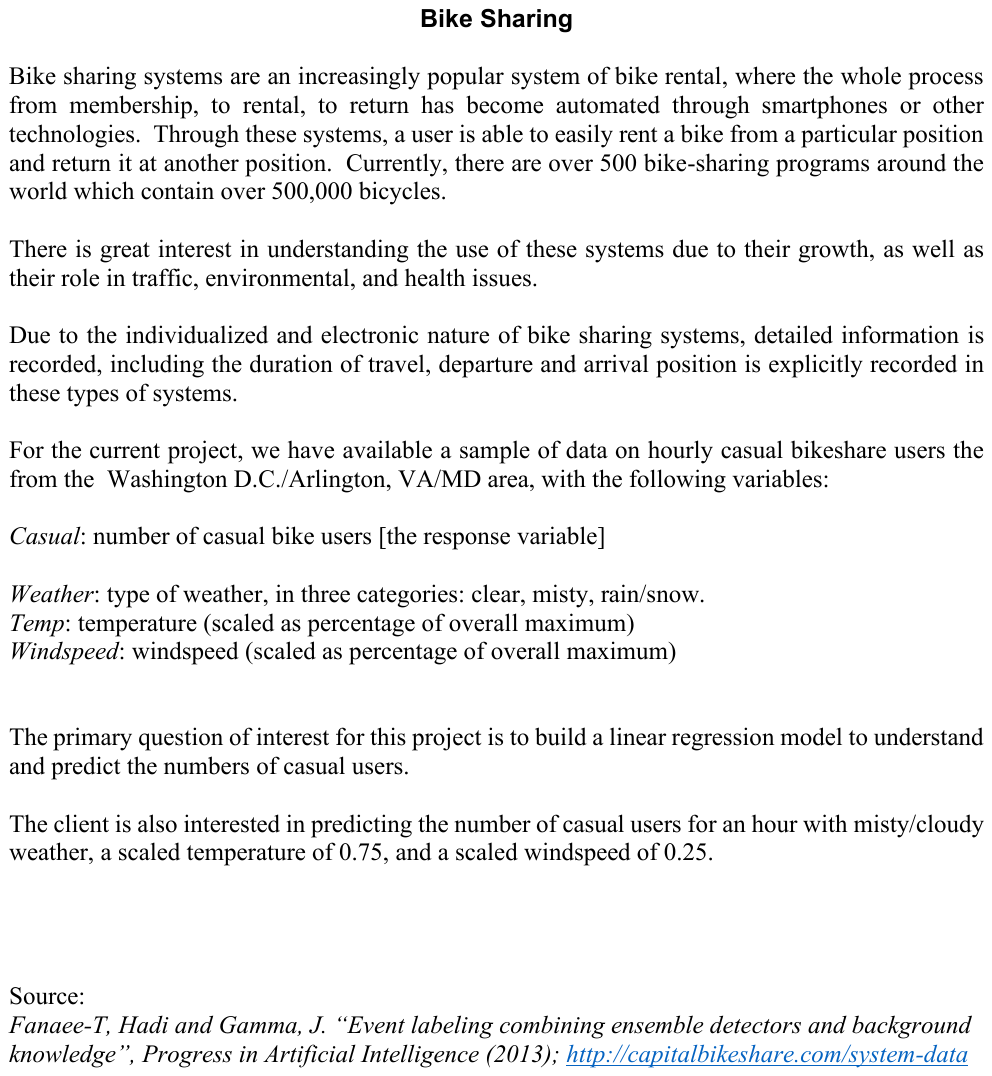}

\end{document}